\documentclass[10pt,journal]{IEEEtran}

\ifCLASSOPTIONcompsoc
  \usepackage[nocompress]{cite}
\else
  \usepackage{cite}
\fi
\ifCLASSINFOpdf
\else
\fi
\usepackage{amsmath}
\usepackage{times}
\usepackage{bm}
\usepackage{amsmath}
\usepackage{amssymb,amsfonts,graphicx,float,multirow,enumitem, url}
\usepackage{bbm}
\usepackage{comment}
\usepackage{placeins}
\usepackage{graphicx,caption} 
\usepackage{blindtext} 
\usepackage{adjustbox} 

\newtheorem{theorem}{Theorem}

\newtheorem{lemma}{Lemma}
\newtheorem{remark}{Remark}
\newtheorem{condition}{Condition}

\allowdisplaybreaks

\newcommand{\pr}{\textsf{pr}} 
\newcommand{\ep}{\textsf{E}} 
\newcommand{\var}{\textsf{var}} 
\newcommand{\cov}{\textsf{cov}} 

\usepackage{color} 
\definecolor{red2}{rgb}{0.7, 0, 0.1}

\usepackage{soul} 

\begin{document}
%
\title{Practical and Powerful Kernel-Based Change-Point Detection}
%
%
%
%

\author{Hoseung~Song
        and~Hao~Chen 

\IEEEcompsocitemizethanks{\IEEEcompsocthanksitem Hoseung Song is with the Public Health Sciences Division at Fred Hutchinson Cancer Research Center, Seattle, WA 98109 USA (e-mail: hsong3@fredhutch.org). \protect \\
Hao Chen is with the Department of Statistics, University of California, Davis, Davis, CA 95616 USA (e-mail: hxchen@ucdavis.edu). 
\IEEEcompsocthanksitem }}
\IEEEtitleabstractindextext{%
\begin{abstract}
Change-point analysis plays a significant role in various fields to reveal discrepancies in distribution in a sequence of observations. While a number of algorithms have been proposed for high-dimensional data, kernel-based methods have not been well explored due to difficulties in controlling false discoveries and mediocre performance. In this paper, we propose a new kernel-based framework that makes use of an important pattern of data in high dimensions to boost power. Analytic approximations to the significance of the new statistics  are derived and fast tests based on the asymptotic results are proposed, offering easy off-the-shelf tools for large datasets. The new tests show superior performance for a wide range of alternatives when compared with other state-of-the-art methods. We illustrate these new approaches through an analysis of a phone-call network data. All proposed methods are implemented in an $\texttt{R}$ package $\texttt{kerSeg}$.
\end{abstract}

\begin{IEEEkeywords}
Kernel methods; Permutation null distribution; General alternatives; Scan statistics; Nonparametrics; High-dimensional data.
\end{IEEEkeywords}}

\maketitle

\IEEEdisplaynontitleabstractindextext

%
\IEEEpeerreviewmaketitle

\ifCLASSOPTIONcompsoc
\IEEEraisesectionheading{\section{Introduction}\label{sec:introduction}}
\else
\section{Introduction}
\label{sec:introduction}
\fi

\IEEEPARstart{R}{ECENT} technological advances have facilitated the collection of high-dimensional data sequences in various high-impact applications, including social sciences \cite{kendrick2018change,wang2017fast}, neuroscience \cite{staudacher2005new, xu2015dynamic}, and computer graphics \cite{radke2005image, tartakovsky2012efficient}. High-dimensional complex data sequences are becoming prevalent and the development of efficient change-point detection method for them is gaining more and more attention. In this paper, we consider the following offline change-point detection problem: given a sequence of independent observations $\{y_{i}\}_{1,\ldots,n}$, $y_{i}\in \mathcal{R}^{d}$, we consider testing the null hypothesis
\begin{equation} \label{eq:H0}
	H_0: \, y_i \sim F_0, \, i = 1, \ldots, n 
\end{equation} 
against the single change-point alternative
\begin{equation} \label{eq:H1} 
	H_1: \exists \, 1 \le \tau < n, \, \, y_i \sim 
	\begin{cases}
		F_0,  \,\,\,\,  i \le \tau \\
		F_1,  \,\,\,\, \text{otherwise}
	\end{cases}
\end{equation}
or the changed interval alternative 
\begin{equation} \label{eq:H2}
	H_2: \exists \, 1 \le \tau_1 < \tau_2 < n, \, \, y_i \sim 
	\begin{cases}
		F_0, \,\,\,\, i = \tau_1 + 1, \hdots, \tau_2 \\
		F_1, \,\,\,\, \text{otherwise}
\end{cases} \end{equation}
where $F_{0}$ and $F_{1}$ are two different disbtributions. 

A number of parametric approaches have been proposed for high-dimensional data, such as the methods in \cite{wang2018high, zhang2021adaptive, jiang2023robust}.  However, parametric approaches for high-dimensional data in general impose strong assumptions that limit their applications. To overcome this, a few nonparametric approaches have been studies, such as the methods using marginal rankings \cite{lung2011homogeneity}, interpoint distances \cite{matteson2014nonparametric, li2020asymptotic}, similarity graphs \cite{chen2015graph, chu2019asymptotic, liu2022fast}, and Fr$\acute{\textrm{e}}$chet mean and variance \cite{dubey2020frechet}.

\subsection{Kernel change-point detection methods and their limitations}

Kernel methods are useful tools under the two-sample hypothesis testing setting for high-dimensional data and they have the potential to capture any types of differences in the distribution.  The most well-known method is the maximum mean discrepancy (MMD) test proposed by \cite{gretton2007kernel} where observations are mapped into a reproducing kernel Hilbert space (RKHS) generated by a given kernel $k(\cdot,\cdot)$ \cite{gretton2009fast, gretton2012kernel, gretton2012optimal}. Compared with kernel methods in the two-sample testing setting, kernel-based change-point analysis received less attention. 

The first practical offline change-point detection method using kernels was proposed by \cite{harchaoui2007retrospective}. They incorporated kernels into dynamic programming algorithms to obtain the optimal location to segment, which is time consuming. Later, a kernel-based test statistic, called the maximum kernel Fisher discriminant ratio, was also proposed by \cite{harchaoui2009kernel}. However, the test relies on the bootstrap resampling method for computing the decision threshold, making the test very slow. \cite{li2015m} proposed MMD-based test statistic by adopting a strategy developed by \cite{zaremba2013b}. This method is computationally efficient, but it does not provide an estimate of the change-point (it provides an estimate of a block of a fixed length that contains the change-point) and requires a large amount of reference data before the change happens. Some other kernel-based change-point detection methods were proposed in \cite{huang2014high, chang2019kernel}, but they do not provide an estimate of the location of change-points when the null hypothesis $H_{0}$ (\ref{eq:H0}) is rejected. Recently, \cite{arlot2019kernel} developed a kernel change-point detection procedure (KCP) that extends the method in \cite{harchaoui2007retrospective}. KCP utilizes a model-selection penalty that allows to select the number of change-points. However, it does not work well under some important types of changes due to the curse of dimensionality (see Section \ref{sec:new} for explanations and  Section \ref{sec:performance} for its performance).  Also, KCP heavily depends on the penalty constant and it is very difficult to control the type I error. Table \ref{tab:simul:tuning} shows the empirical size of KCP under different dimensions and penalty constants for Gaussian data when $n=200$. We see that the empirical size of the test is sensitive to the penalty constant, particularly for high-dimensional data. 
\begin{table}[h]
	\caption{Empirical size of KCP under different dimensions and penalty constants for Gaussian data}
	\vspace{0.5em}
	\label{tab:simul:tuning}
	\centering
	\begin{tabular}{c|c|ccc}
		\hline
		\multirow{2}{*}{$d=100$} &  Penalty constant & 0.345 & 0.340 & 0.335  \\
		& Empirical sizes &  0.041 & 0.056 & 0.084  \\
		\hline
		\multirow{2}{*}{$d=500$} & Penalty constant &  0.0590 & 0.0585 & 0.0580  \\
		&  Empirical sizes &  0.028 & 0.051 & 0.081    \\
		\hline
		\multirow{2}{*}{$d=1000$} &  Penalty constant &  0.0287 & 0.0282 & 0.0277  \\
		&  Empirical sizes &  0.009 & 0.036 & 0.159 \\
		\hline
	\end{tabular}
\end{table}

\subsection{Our contribution}

To the best of our knowledge, all existing kernel change-point detection methods are restricted to specific types of alternatives and miss some important types of changes, such as location and scale changes. We propose new kernel-based test statistics that perform well for a wide range of alternatives and achieves high power in detecting and estimating change-points in the high-dimensional sequence compared to other state-of-the-art change-point detection methods.  The new methods are easy to implement and have no tuning parameter.  We also propose fast tests and derive analytic formulas for type I error control, allowing instant application to large datasets. The new methods are implemented in a $\texttt{R}$ package $\texttt{kerSeg}$.

The organization of the paper is as follows. In Section \ref{sec:new}, we propose new scan statistics for the single change-point and changed-interval alternatives. The asymptotic behavior of the new test statistics, the analytical $p$-value approximations, and fast tests are provided in Section \ref{sec:theory}. Section \ref{sec:performance} examines the performance of the new tests under various simulation settings. The new approaches are illustrated by a real data application on a phone-call network data in Section \ref{sec:real}. We conclude with discussion in Section \ref{sec:conclusion}.


\section{New scan statistics} \label{sec:new}

Since there is no distributional assumption, we work under the permutation null distribution, which places $1/n!$ probability on each of the $n!$ permutations of $\{y_{i}\}_{1,\ldots,n}$. We use $\pr$, $\ep$, $\var$, and $\cov$ to denote the probability, expectation, variance, and covaraince, repectively, under the permutation null distribution. In addition, without further specification, we use the Gaussian kernel with the median heuristic as the bandwidth parameter. 

\subsection{Scan statistics for the single change-point alternative} \label{subsec:single}

\begin{figure}[h]
	\centering
	\includegraphics[width=\columnwidth]{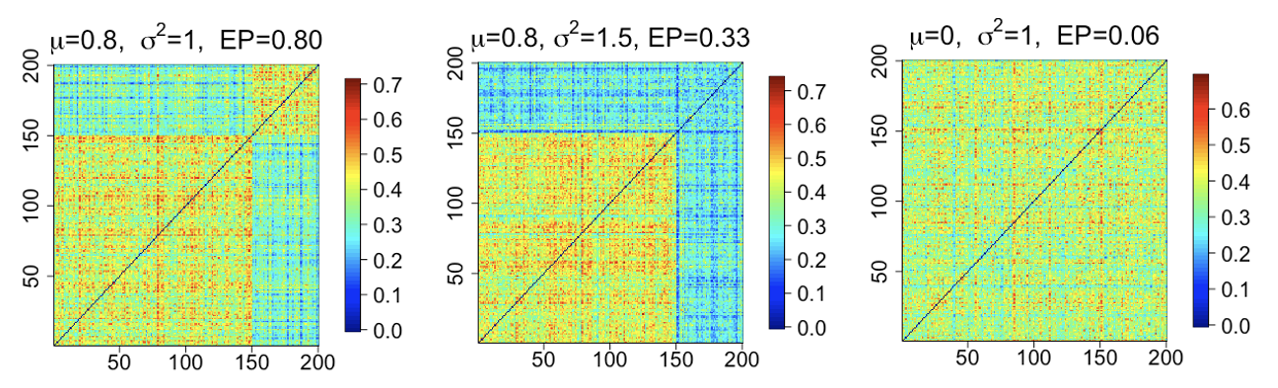}
	\caption{Heatmaps of kernel matrices under different cases, and the estimated power of $\max_{n_{0}\le t \le n_{1}}\textrm{MMD}^2_{u}(t)$ (denoted by `EP' in the title).}
	\label{fig:motivation}
\end{figure}

In the above mentioned kernel change-point methods \cite{harchaoui2007retrospective,huang2014high, li2015m}, the MMD-based test statistic was used for constructing the scan statistics. 
MMD-based tests were proved to be consistent against all alternatives for the two-sample testing \cite{gretton2012kernel}. However, it could have very low power under finitie sample sizes, such as in hundreds or thousands, for some common alternatives \cite{song2020generalized}. The same problem also occurs under the change-point setting. For example, we consider Gaussian data $N_{d}(\textbf{0}_{d}, \Sigma)$ vs $N_{d}(\mu\textbf{1}_{d}, \sigma^2\Sigma)$ where $n=200$, $\tau = 150$, $\Sigma_{i,j} = 0.4^{|i-j|}$, $\textbf{0}_{d}$ and $\textbf{1}_{d}$ are $d$ dimensional vectors of zeros, and ones, respectively, and $d=50$. Based on an unbiased estimator of $\textrm{MMD}^2$ \cite{gretton2007kernel}, its scan statistic can be computed as
\begin{align*}
	\textrm{MMD}^2_{u}(t) &= \frac{1}{t(t-1)}\sum_{i=1}^{t}\sum_{j=1,j\ne i}^{t}k(y_{i},y_{j}) \\
	& + \frac{1}{(n-t)(n-t-1)}\sum_{i=t+1}^{n}\sum_{j=t+1,j\ne i}^{n}k(y_{i},y_{j}) \\
	& - \frac{2}{t(n-t)}\sum_{i=1}^{t}\sum_{j=t+1}^{n}k(y_{i},y_{j}) \\
	& \stackrel{\triangle}{=} \alpha(t) + \beta(t) - 2\gamma(t) \\
	& = \big\{\alpha(t) - \gamma(t)\big\} + \big\{\beta(t) - \gamma(t)\big\}.
\end{align*}
Figure \ref{fig:motivation} shows heatmaps of kernel matrices under different cases and the estimated power of $\max_{n_{0}\le t \le n_{1}}\textrm{MMD}^2_{u}(t)$ where $n_{0}=0.05n$ and $n_{1}=n-n_{0}$ by 100 trials based on 10,000 bootstrap replicates. Under location changes (left panel), since kernel values are proportional to the similarity between two observations, we would expect both $\alpha(t)$ and $\beta(t)$ to be larger than $\gamma(t)$, which leads to large $\textrm{MMD}^2_{u}(t)$ and high power. However, when there are additional variance changes (middle panel), due to the curse of dimensionality, samples from the distribution with a larger variance could be closer to samples from the distribution with a smaller variance (see more discussions on this phenomenon in \cite{chen2017new}). Then the effects of $\alpha(t)-\gamma(t)$ and $\beta(t)-\gamma(t)$ could offset, which results in lower power with additional variance change on top of the mean change. 

Let $g_{i}(t) = I_{i>t}$, where $I_{A}$ is an indicator function that takes value 1 if $A$ is true and 0 otherwise, and $k_{ij} = k(y_{i},y_{j})$ $(i,j=1,\ldots,n)$. The quantities $\alpha(t)$ and $\beta(t)$ can be written as
\begin{align*}
	\alpha(t) &= \frac{1}{t(t-1)}\sum_{i=1}^{n}\sum_{j=1, j\ne i}^{n}k_{ij}I_{g_{i}(t) = g_{j}(t) = 0}, \\
	\beta(t) &= \frac{1}{(n-t)(n-t-1)}\sum_{i=1}^{n}\sum_{j=1, j\ne i}^{n}k_{ij}I_{g_{i}(t) = g_{j}(t) = 1}.
\end{align*}
In light of the work in \cite{song2020generalized}, we define a scan statistic to aggregate deviations of $\alpha(t)$ and $\beta(t)$ from their expectations under the permutation null distribution in both directions:
\begin{align}
	\textrm{GKCP}(t) &= \big(\alpha(t) - \ep(\alpha(t)), \beta(t)-\ep(\beta(t))\big) \notag \\
	&\ \ \ \ \ \ \  \times \Sigma_{t}^{-1}\left(\begin{tabular}{c}
		$\alpha(t) - \ep(\alpha(t))$ \\
		$\beta(t)-\ep(\beta(t))$
	\end{tabular}\right), 
\end{align}
where $\Sigma_{t} = \var((\alpha(t),\beta(t))^{T})$. Under the permutation null distribution, the analytic expressions for the expectation and the variance of $\alpha(t)$ and $\beta(t)$ can be calculated through combinatorial analysis, similarly as in \cite{song2020generalized}. They are provided in Lemma \ref{expectation} (proof in Supplement A).
\begin{lemma} \label{expectation}
	Under the permutation null, we have
	\begin{align*}
		\ep\left(\alpha(t)\right) &= \ep\left(\beta(t)\right) = \frac{1}{n(n-1)}R_{0} \stackrel{\Delta}{=} \bar{k}, \\
		\var\left(\alpha(t)\right) &= \frac{2R_{1}p_{1}(t) + 4R_{2}p_{2}(t) + R_{3}p_{3}(t)}{t^2(t-1)^2} - \bar{k}^2, \\
		\var\left(\beta(t)\right) &= \frac{2R_{1}q_{1}(t) + 4R_{2}q_{2}(t) + R_{3}q_{3}(t)}{(n-t)^2(n-t-1)^2} - \bar{k}^2, \\
		\cov\left(\alpha(t), \beta(t)\right) &= \frac{R_{3}}{n(n-1)(n-2)(n-3)} - \bar{k}^2,
	\end{align*}
	where
	\begin{align*}
		R_{0} &= \sum_{i=1}^{n}\sum_{j=1,j\ne i}^{n}k_{ij}, \ \  \ R_{1} = \sum_{i=1}^{n}\sum_{j=1,j\ne i}^{n}k^2_{ij}, \\
		R_{2} &= \sum_{i=1}^{n}\sum_{j=1,j\ne i}^{n}\sum_{u=1,u\ne j, u\ne i}^{n}k_{ij}k_{iu}, \\ 
		R_{3} &= \sum_{i=1}^{n}\sum_{j=1,j\ne i}^{n}\sum_{u=1,u\ne j, u\ne i}^{n}\sum_{v=1,v\ne u, v\ne j, v\ne i}^{n}k_{ij}k_{uv},
	\end{align*}
	\begin{align*}
		&p_{1}(t) = \frac{t(t-1)}{n(n-1)}, \ \ \ p_{2}(t) = p_{1}(t)\frac{t-2}{n-2}, \\ 
		&p_{3}(t) = p_{2}(t)\frac{t-3}{n-3}, \\
		&q_{1}(t) = \frac{(n-t)(n-t-1)}{n(n-1)}, \ \ \ q_{2}(t) = q_{1}(t)\frac{n-t-2}{n-2}, \\
		&q_{3}(t) = q_{2}(t)\frac{n-t-3}{n-3}.
	\end{align*}
\end{lemma}

To test $H_{0}$ (\ref{eq:H0}) versus $H_{1}$ (\ref{eq:H1}), the following scan statistic is used:
\begin{align}
	\max_{n_{0}\le t \le n_{1}}\textrm{GKCP}(t), 
\end{align}
where $n_{0}$ and $n_{1}$ are pre-specified constraints on the region where the change-point $\tau$ is searched. By default, we can set $n_{0} = [0.05n]$\footnote{$[x]$ denotes the largest integer that is no larger than $x$.} and $n_{1} = n-n_{0}$. If there are prior information on the range of the potential change-point, then $n_{0}$ and $n_{1}$ can be specified accordingly. The null hypothesis $H_{0}$ (\ref{eq:H0}) is rejected if the scan statistic is greater than a threshold. Explorations on how to choose the threshold to control the type I error are discussed in Section \ref{sec:theory}.

\subsection{Scan statistics for the changed-interval alternative} \label{subsec:interval}

Here, we define the test statistic for testing $H_{0}$ (\ref{eq:H0}) against the changed-interval alternative $H_{2}$ (\ref{eq:H2}). Similar to the singe change-point alternative, each possible interval $(t_{1}, t_{2}]$ divides the data sequence into two groups. Then, for any candidate interval $(t_{1}, t_{2}]$, the test statistics $Z_{D}(t_{1},t_{2})$ and $Z_{W}(t_{1},t_{2})$ can be defined in a similar manner to the single change-point alternative. Under the permutation null, the analytic expression for $\ep(D(t_{1},t_{2}))$, $\ep(W(t_{1},t_{2}))$, $\var(D(t_{1},t_{2}))$, and $\var(W(t_{1},t_{2}))$ can be obtained similary as in the single change-point setting. The scan statistic involves a maximization over $t_{1}$ and $t_{2}$, 
\begin{align}
	\max_{\substack{1<t_{1}<t_{2}\le n \\ n_{0}\le t_{2}-t_{1} \le n_{1}}}\textrm{GKCP}(t_{1},t_{2}),
\end{align}
where $n_{0}$ and $n_{1}$ are constraints on the window size.


\section{Analytical $p$-value approximations and fast tests} \label{sec:theory}

Given the test statistic, the next step is to determine how large the test statistic needs to provide sufficient evidence to reject the null hypothesis of homogeneity. That is, we are concerned with the tail probabilities of the scan statistic under $H_{0}$ (\ref{eq:H0}),
\begin{equation}  \label{tail:single}
	\pr\left(\max_{n_{0}\le t \le n_{1}}\textrm{GKCP}(t) > b\right) 
\end{equation}
for the single change-point alternative, and
\begin{equation}  \label{tail:interval}
	\pr\left(\max_{\substack{1<t_{1}<t_{2}\le n \\ n_{0}\le t_{2}-t_{1} \le n_{1}}}\textrm{GKCP}(t_{1},t_{2}) > b\right) 
\end{equation}
for the changed-interval alternative. The threshold can be approximated by drawing random permutations of the sequence, which is time consuming. We thus investigate the stochastic process to see if there is any way to make the test faster.

Based on the results in \cite{song2020generalized}, it can be shown that
\begin{equation}
	\textrm{GKCP}(t) = Z_{D}^2(t) + Z_{W}^2(t),
\end{equation}
where
\begin{align}
	Z_{D}(t) = \frac{D(t)-\ep(D(t))}{\sqrt{\var(D(t))}}, \ \ \ Z_{W}(t) = \frac{W(t)-\ep(W(t))}{\sqrt{\var(W(t))}}
\end{align}
with 
\begin{align}
	D(t) &= t(t-1)\alpha(t) - (n-t)(n-t-1)\beta(t), \\
	W(t) &= \frac{n-t}{n}\alpha(t) + \frac{t}{n}\beta(t).
\end{align}
\cite{song2020generalized} showed that the test of the two-sample version of $Z_{W}(t)$ is equivalent to the test based on $\textrm{MMD}^2_{u}$ under the permutation null, but the limiting distribution of $\textrm{MMD}^2_{u}$ is not easy to handle \cite{gretton2007kernel}. Due to the intrinsic relation between the test based on $\textrm{MMD}^2_{u}$ and $W(t)$, it is also not easy to handle the limiting distribution of $W(t)$. Hence, we first define a related quantity, an weighted version of $W(t)$, to obtain the tractable asymptotic results:
\begin{align}
	W_{r}(t) = r\frac{n-t}{n}t(t-1)\alpha(t) + \frac{t}{n}(n-t)(n-t-1)\beta(t).
\end{align}
$Z_{W,r}(t)$ is the standardized $W_{r}(t)$, where $r$ is a constant. 

In the rest of this chapter, we first study the asymptotic properties of the stochastic processes $\{Z_{D}([nu]): 0<u<1\}$, $\{Z_{W,r}([nu]): 0<u<1\}$, $\{Z_{D}([nu],[nv]): 0<u<v<1\}$, and $\{Z_{W,r}([nu],[nv]): 0<u<v<1\}$ (Section \ref{subsec:asymp}). We then derive analytic approximations to the tail probabilities under the Gaussian field approximation (Section \ref{subsec:pval}). We improve our approximations by correcting the skewness in the marginal distributions (Section \ref{subsec:skewness}) and these approximations are checked by numerical studies in Section \ref{subsec:check}. Finally, we propose fast tests based on the asymptotic results in Section \ref{subsec:fast}.

\subsection{Asymptotic distributions of the basic processes} \label{subsec:asymp}

In this section, we derive the limiting distributions of $\{Z_{D}([nu]): 0<u<1\}$ and $\{Z_{W,r}([nu]): 0<u<1\}$ for the single change-point alternative and $\{Z_{D}([nu],[nv]): 0<u<v<1\}$ and $\{Z_{W,r}([nu],[nv]): 0<u<v<1\}$ for the changed-interval alternative.

In the following, we write $a_n = O(b_n)$ when $a_n$ has the same order as $b_n$ and $a_n = o(b_n)$ when $a_n$ is dominated by $b_n$ asymptotically, i.e., $\lim_{n\rightarrow\infty}(a_{n}/b_{n}) = 0$. Let $\tilde{k}_{ij} = (k_{ij} - \bar{k})I_{i\ne j}$ and $\tilde{k}_{i\cdot} = \sum_{j=1,j\ne i}^{n}\tilde{k}_{ij}$ for $i=1,\ldots,n$. We work under the following two conditions.
\begin{condition} \label{condition1}
	$\sum_{i=1}^{n}|\tilde{k}_{i\cdot}|^s = o\Big(\left\{\sum_{i=1}^{n}\tilde{k}_{i\cdot}^2\right\}^{s/2}\Big)$ for all integers $s>2$.
\end{condition}
\begin{condition} \label{condition2}
	$\sum_{i=1}^{n}\sum_{j=1,j\ne i}^{n}\tilde{k}_{ij}^2 = o\Big(\sum_{i=1}^{n}\tilde{k}_{i\cdot}^2\Big)$.
\end{condition}
\begin{theorem} \label{asymp1}
	Under Condition \ref{condition1} and \ref{condition2}, as $n\rightarrow\infty$,
	\begin{enumerate}
		\item $\{Z_{D}([nu]): 0<u<1\}$ converges to a Gaussian process in finite dimensional distributions, which we denote as $\{Z_{D}^{*}(u): 0<u<1\}$.
		\item $\{Z_{D}([nu],[nv]): 0<u<v<1\}$ converges to a two-dimensional Gaussian random field in finite dimensional distributions, which we denote as $\{Z_{D}^{*}(u,v): 0<u<v<1\}$.
		\item $\{Z_{W,r}([nu]): 0<u<1\}$ converges to a Gaussian process in finite dimensional distributions when $r\ne1$, which we denote as $\{Z_{W,r}^{*}(u): 0<u<1\}$.
		\item $\{Z_{W,r}([nu],[nv]): 0<u<v<1\}$ converges to a two-dimensional Gaussian random field in finite dimensional distributions $r\ne1$, which we denote as $\{Z_{W,r}^{*}(u,v): 0<u<v<1\}$.
	\end{enumerate}
\end{theorem} 
The proof for this theorem is in Supplement B.

\begin{remark}
	Condition \ref{condition1} can be satisfied when $|\tilde{k}_{i\cdot}| = O(n^{\delta})$ for a constant $\delta$, $\forall i$, and Condition \ref{condition2} would further be satiesfied if we also have $\tilde{k}_{ij} = O(n^{\kappa})$ for a constant $\kappa < \delta - 0.5$, $\forall i,j$. When there is no big outlier in the data, it is not hard to have all these conditions satisfied when one uses the Gaussian kernel with the median heuristic.
\end{remark}

Let $\rho_{D}^{*}(u,v) = \cov\left(Z_{D}^{*}(u), Z_{D}^{*}(v)\right)$ and $\rho_{W,r}^{*}(u,v) = \cov\big(Z_{W,r}^{*}(u), Z_{W,r}^{*}(v)\big)$. The explicit covariance functions of the limiting Gaussian processes, $\{Z_{D}^{*}(u): 0<u<1\}$ and $\{Z_{W,r}^{*}(u): 0<u<1\}$ are stated in the following theorem.

\begin{theorem} \label{thm:explicit}
	The exact expression for $\rho_{D}^{*}(u,v)$ and $\rho_{W,r}^{*}(u,v)$ are
	\begin{align*}
		&\rho_{D}^{*}(u,v) = \frac{(u\wedge v)\left(1-(u\vee v)\right)}{\sqrt{u(1-u)v(1-u)}}, \\
		&\rho_{W,r}^{*}(u,v) = \frac{2R_{1}\{r^2(u\wedge v)\left(1-(u\wedge v)\right)\left(1-(u\vee v)^2\right) \}}{(u\vee v)\left(1-(u\wedge v)\right)\sigma_{W,r}^{*}(u)\sigma_{W,r}^{*}(v)} \\
		&+ \frac{2R_{1}\{r\left((u\vee v)-1\right)\left(3uv-(u\wedge v)^2(2(u\vee v)+1)\right)\}}{(u\vee v)\left(1-(u\wedge v)\right)\sigma_{W,r}^{*}(u)\sigma_{W,r}^{*}(v)} \\
		&+ \frac{2R_{1}\{uv\left(2-(u\wedge v)\right)\left(1-(u\vee v)\right)\}}{(u\vee v)\left(1-(u\wedge v)\right)\sigma_{W,r}^{*}(u)\sigma_{W,r}^{*}(v)} \\
		&+\frac{4R_{2}uv(1-u)(1-v)(r-1)^2}{(u\vee v)\left(1-(u\wedge v)\right)\sigma_{W,r}^{*}(u)\sigma_{W,r}^{*}(v)},
	\end{align*}
	where $u\wedge v = \min(u,v)$, $u\vee v = \max(u,v)$, and
	\begin{align*}
		&\sigma_{W,r}^{*}(u) \\
		&= \sqrt{2R_{1}\{r(1-u)+u\}^2 + (4R_{1}+4R_{2})u(1-u)(r-1)^2}.
	\end{align*}
\end{theorem}
The above theorem is proved through combinatorial analysis and the details are in Supplement C. From the above theorem, we see that the limiting process $\{Z_{D}^{*}(u): 0<u<1\}$ does not depend on kernel values, while $\{Z_{W,r}^{*}(u): 0<u<1\}$ depends on kernel values.

\subsection{Asymptotic $p$-value approximations} \label{subsec:pval}

We now examine the asymptotic behavior of tail probabilities (\ref{tail:single}) and (\ref{tail:interval}). Following similar arguments in the proof for Proposition 3.4 in \cite{chen2015graph}, when $n_{0}, n_{1}, n, b \rightarrow \infty$ in a way such that for some $0 < x_{0} < x_{1} < 1$ and $b_{0} > 0$, $n_{0}/n \rightarrow x_{0}$, $n_{1}/n \rightarrow x_{1}$, and $b/\sqrt{n} \rightarrow b_{0}$, as $n\rightarrow \infty$, we have
\begin{align}
	&\pr\left(\max_{n_{0}\leq t \leq n_{1}}|Z_{D}^{*}(t/n)| > b\right)  \notag \\
	&\sim 2b\phi(b)\int_{x_{0}}^{x_{1}}h_{D}^{*}(x)\nu\Big(b_{0}\sqrt{2h_{D}^{*}(x)}\Big)dx, \label{pvalapprox1}\\
	&\pr\left(\max_{n_{0}\leq t_{2}-t_{1} \leq n_{1}}|Z_{D}^{*}(t_{1}/n, t_{2}/n)| > b\right) \notag \\
	&\sim 2b^3\phi(b)\int_{x_{0}}^{x_{1}}\left(h_{D}^{*}(x)\nu\Big(b_{0}\sqrt{2h_{D}^{*}(x)}\Big)\right)^2(1-x)dx,  \\
	&\pr\left(\max_{n_{0}\leq t \leq n_{1}}Z_{W,r}^{*}(t/n) > b\right)  \notag \\
	&\sim b\phi(b)\int_{x_{0}}^{x_{1}}h_{W,r}^{*}(x)\nu\Big(b_{0}\sqrt{2h_{W,r}^{*}(x)}\Big)dx, \\
	&\pr\left(\max_{n_{0}\leq t_{2}-t_{1} \leq n_{1}}Z_{W,r}^{*}(t_{1}/n, t_{2}/n) > b\right) \notag \\
	&\sim b^3\phi(b)\int_{x_{0}}^{x_{1}}\left(h_{W,r}^{*}(x)\nu\Big(b_{0}\sqrt{2h_{W,r}^{*}(x)}\Big)\right)^2(1-x)dx, \label{pvalapprox2}
\end{align}
where the function $\nu(\cdot)$ can be numerically estimated as
\begin{equation*}
	\nu(s) \approx \frac{(2/s)\left(\Phi(s/2)-0.5\right)}{(s/2)\Phi(s/2)+\phi(s/2)}
\end{equation*}
according to \cite{siegmund2007statistics} with $\Phi(\cdot)$ and $\phi(\cdot)$ being the standard normal cumulative density function and probability density function, respectively, and
\begin{align*}
	h_{D}^{*}(x) &= \lim_{s\nearrow x}\frac{\partial\rho_{D}^{*}(s,x)}{\partial s}  = -\lim_{s\searrow x}\frac{\partial\rho_{D}^{*}(s,x)}{\partial s}, \\
	h_{W,r}^{*}(x) &= \lim_{s\nearrow x}\frac{\partial\rho_{W,r}^{*}(s,x)}{\partial s} = -\lim_{s\searrow x}\frac{\partial\rho_{W,r}^{*}(s,x)}{\partial s}.
\end{align*}

\begin{remark}
	In practice, when using (\ref{pvalapprox1})--(\ref{pvalapprox2}) for finite sample, we use
	\begin{align*}
		&\pr\left(\max_{n_{0}\leq t \leq n_{1}}|Z_{D}(t)| > b\right)  \\
		&\sim 2b\phi(b)\sum_{n_{0}\le t \le n_{1}}C_{D}(t)\nu\Big(b\sqrt{2C_{D}(t)}\Big), \\
		&\pr\left(\max_{n_{0}\leq t_{2}-t_{1} \leq n_{1}}|Z_{D}(t_{1}, t_{2})| > b\right) \\
		&\sim 2b^3\phi(b)\sum_{n_{0}\le t \le n_{1}}\left(C_{D}(t)\nu\Big(b\sqrt{2C_{D}(t)}\Big)\right)^2(n-t),  \\
		&\pr\left(\max_{n_{0}\leq t \leq n_{1}}Z_{W,r}(t) > b\right) \\
		&\sim b\phi(b)\sum_{n_{0}\le t \le n_{1}}C_{W,r}(t)\nu\Big(b\sqrt{2C_{W,r}(t)}\Big), \\
		&\pr\left(\max_{n_{0}\leq t_{2}-t_{1} \leq n_{1}}Z_{W,r}(t_{1}, t_{2}) > b\right) \\
		&\sim b^3\phi(b)\sum_{n_{0}\le t \le n_{1}}\left(C_{W,r}(t)\nu\Big(b\sqrt{2C_{W,r}(t)}\Big)\right)^2(n-t),  
	\end{align*}
	where
	\begin{align*}
		C_{D}(t) = \frac{\partial\rho_{D}(s,t)}{\partial s}\Bigr|_{s=t}, \ \ \ C_{W,r}(t) = \frac{\partial\rho_{W,r}(s,t)}{\partial s}\Bigr|_{s=t}
	\end{align*}
	with $\rho_{D}(u,v) = \cov\left(Z_{D}(u), Z_{D}(v)\right)$ and $\rho_{W,r}(u,v) = \cov\left(Z_{W,r}(u), Z_{W,r}(v)\right)$. The explicit expressions for $C_{D}(t)$ and $C_{W,r}(t)$ can be calculated in the similar manner to the proof of Theorem \ref{thm:explicit} and they are provided in Supplement C.
\end{remark}

\subsection{Skewness correction} \label{subsec:skewness}

The analytical $p$-value approximations based on the asymptotic results provide a practical tool for large datasets. However, they become less precise if we set $n_{0}$ and $n_{1}$ close to the two ends since the convergence of $Z_{D}(t)$ and $Z_{W,r}(t)$ to the Gaussian process is slow as $t/n$ is close to 0 or 1. Hence, we improve the accuracy of the analytical $p$-value approximations for finite sample sizes by skewness correction. As the skewness depends on the value of $t$, we adopt a similar treatment discussed in \cite{chen2015graph} and we add extra terms in the analytic formulas to correct skewness. 

After skewness correction, the analytical $p$-value approximations are
\begin{align}
	&\pr\left(\max_{n_{0}\leq t \leq n_{1}}|Z_{D}^{*}(t/n)| > b\right)  \notag \\
	&\sim 2b\phi(b)\int_{x_{0}}^{x_{1}}S_{D}(x)h_{D}^{*}(x)\nu\Big(b_{0}\sqrt{2h_{D}^{*}(x)}\Big)dx, \\
	&\pr\left(\max_{n_{0}\leq t_{2}-t_{1} \leq n_{1}}|Z_{D}^{*}(t_{1}/n, t_{2}/n)| > b\right) \notag \\
	&\sim 2b^3\phi(b)\int_{x_{0}}^{x_{1}}S_{D}(x)\left(h_{D}^{*}(x)\nu\Big(b_{0}\sqrt{2h_{D}^{*}(x)}\Big)\right)^2(1-x)dx,  \\
	&\pr\left(\max_{n_{0}\leq t \leq n_{1}}Z_{W,r}^{*}(t/n) > b\right) \notag \\
	&\sim b\phi(b)\int_{x_{0}}^{x_{1}}S_{W,r}(x)h_{W,r}^{*}(x)\nu\Big(b_{0}\sqrt{2h_{W,r}^{*}(x)}\Big)dx, \\
	&\pr\left(\max_{n_{0}\leq t_{2}-t_{1} \leq n_{1}}Z_{W,r}^{*}(t_{1}/n, t_{2}/n) > b\right) \notag \\
	&\sim b^3\phi(b) \notag \\
	&\times \int_{x_{0}}^{x_{1}}S_{W,r}(x)\left(h_{W,r}^{*}(x)\nu\Big(b_{0}\sqrt{2h_{W,r}^{*}(x)}\Big)\right)^2(1-x)dx, 
\end{align}
where
\begin{align*}
	S_{D}(t) &= \frac{\exp\big\{\frac{1}{2}(b-\hat{\theta}_{b,D}(t))^2+\frac{1}{6}\gamma_{D}(t)\hat{\theta}_{b,D}^{3}(t)\big\}}{\sqrt{1+\gamma_{D}(t)\hat{\theta}_{b,D}(t)}}, \\
	S_{W,r}(t) &= \frac{\exp\big\{\frac{1}{2}(b-\hat{\theta}_{b,W,r}(t))^2+\frac{1}{6}\gamma_{W,r}(t)\hat{\theta}_{b,W,r}^{3}(t)\big\}}{\sqrt{1+\gamma_{W,r}(t)\hat{\theta}_{b,W,r}(t)}}, 
\end{align*}
with
\begin{align*}
	\hat{\theta}_{b,D}(t) &= \frac{\sqrt{1+2\gamma_{D}(t)b}-1}{\gamma_{D}(t)}, \ \ \ \gamma_{D}(t) = \ep\left(Z_{D}^3(t)\right), \\ 
	\hat{\theta}_{b,W,r}(t) &= \frac{\sqrt{1+2\gamma_{W,r}(t)b}-1}{\gamma_{W,r}(t)}, \ \ \  \gamma_{W,r}(t) = \ep\left(Z_{W,r}^3(t)\right).
\end{align*}
To obtain $\gamma_{D}(t)$ and $\gamma_{W,r}(t)$, we need to figure out $\ep\left(D^3(t)\right)$ and $\ep\left(W_{r}^3(t)\right)$. The exact analytic expressions of $\ep\left(D^3(t)\right)$ and $\ep\left(W_{r}^3(t)\right)$ are complicated and they are provided in Supplement D.

\begin{remark}
	When the marginal distribution is highly left-skewed, the skewness is so small that $1+2\gamma(t)b$ could be negative. Since this problem usually happens when $t/n$ is close to 0 or 1, we apply a heuristic fix discussed in \cite{chen2015graph} by extrapolating $\hat{\theta}_{b}(t)$.
\end{remark}

\subsection{Checking $p$-value approximations under finite $n$} \label{subsec:check}

In this section, we check how the analytical $p$-value approximations work for finite samples. To this end, we compare the critical values for 0.05 $p$-value threshold obtained from doing 10,000 permutations and the critical values obtained in Section \ref{subsec:pval} and \ref{subsec:skewness} under various simulation settings. Here, we focus on the single-change-point alternative.

We consider three distributions (multivariate Gaussian (C1), multivariate $t_{5}$ (C2), multivariate log-normal (C3)) under various dimensions ($d=100, 500, 1000$). In each simulation, two randomly simulated $n=1,000$ sequences are generated. The analytic approximations depend on constraints $n_{0}$ and $n_{1}$. To make things simple, we set $n_{1} = n - n_{0}$.

Since the asymptotic $p$-value approximation of $Z_{D}(t)$ without skewness correction does not depend on kernel values, the critical value is determined by $n$, $n_{0}$, and $n_{1}$ only. On the other hand, the asymptotic $p$-value approximation of $Z_{W,r}(t)$ without skewness correction and the skewness corrected $p$-value approximations of $Z_{D}(t)$ and $Z_{W,r}(t)$ depend on certain kenel values. 

\begin{table}[h!]
	\centering 
	\caption{Critical values for the single change-point scan statistic $\max_{n_0 \le t \le n_1} Z_{D}(t)$ at 0.05 significance level}
	\label{tab:zdcv}
	\begin{tabular}{c|cc|cc|cc|cc} 
		\hline  
		& \multicolumn{2}{c|}{$n_0 = 100$} &  \multicolumn{2}{c|}{$n_0 = 75$} &  \multicolumn{2}{c|}{$n_0 = 50$} &  \multicolumn{2}{c}{$n_0 = 25$} \\ 
		\hline
		& Ana & Per &  Ana &Per & Ana &Per &  Ana &Per  \\
		\hline 
		Gaussian & 3.00 & 3.01 & 3.05 & 3.04 & 3.10 & 3.09 & 3.16 & 3.14  \\
		$d = 100$ & 3.00 & 3.01 & 3.05 & 3.03 & 3.10 & 3.11 & 3.16 & 3.15 \\ 
		Gaussian & 3.00 & 3.01 & 3.05 & 3.04 & 3.10 & 3.10 & 3.16 & 3.16  \\
		$d = 500$ & 3.00 & 3.01 & 3.05 & 3.05 & 3.10 & 3.10 & 3.16 & 3.16 \\ 
		Gaussian & 3.00 & 3.01 & 3.05 & 3.04 & 3.10 & 3.10 & 3.16 & 3.14   \\
		$d = 1000$ & 3.00 & 2.99 & 3.05 & 3.06 & 3.10 & 3.10 & 3.16 & 3.15 \\ 
		\hline
		MV-$t_{5}$ & 3.00 & 3.02 & 3.05 & 3.03 & 3.10 & 3.10 & 3.16 & 3.16  \\
		$d = 100$ & 3.00 & 3.00 & 3.05 & 3.04 & 3.10 & 3.10 & 3.16 & 3.16 \\ 
		MV-$t_{5}$ & 3.00 & 2.99 & 3.04 & 3.04 & 3.10 & 3.09 & 3.16 & 3.16  \\
		$d = 500$ & 3.00 & 2.99 & 3.04 & 3.03 & 3.10 & 3.09 & 3.16 & 3.16 \\ 
		MV-$t_{5}$ & 3.00 & 2.99 & 3.05 & 3.04 & 3.10 & 3.08 & 3.17 & 3.18  \\
		$d = 1000$ & 3.00 & 2.99 & 3.05 & 3.05 & 3.10 & 3.09 & 3.17 & 3.16 \\
		\hline
		Log-normal & 3.00 & 2.98 & 3.05 & 3.02 & 3.10 & 3.08 & 3.16 & 3.16  \\
		$d = 100$ & 3.00 & 2.99 & 3.05 & 3.04 & 3.10 & 3.04 & 3.16 & 3.15 \\
		Log-normal & 3.00 & 3.00 & 3.04 & 3.04 & 3.10 & 3.09 & 3.16 & 3.16  \\
		$d = 500$ & 3.00 & 2.99 & 3.04 & 3.03 & 3.10 & 3.09 & 3.16 & 3.16 \\ 
		Log-normal & 3.00 & 2.99 & 3.05 & 3.06 & 3.10 & 3.07 & 3.17 & 3.15  \\
		$d = 1000$ & 3.00 & 2.99 & 3.05 & 3.02 & 3.10 & 3.09 & 3.17 & 3.14 \\
		\hline 
	\end{tabular} 
\end{table} 
\begin{table}[h!]
	\centering 
	\caption{Critical values for the single change-point scan statistic $\max_{n_0 \le t \le n_1} Z_{W,1.2}(t)$ at 0.05 significance level}
	\label{tab:zwcv1}
	\begin{tabular}{c|cc|cc|cc|cc} 
		\hline  
		& \multicolumn{2}{c|}{$n_0 = 100$} &  \multicolumn{2}{c|}{$n_0 = 75$} &  \multicolumn{2}{c|}{$n_0 = 50$} &  \multicolumn{2}{c}{$n_0 = 25$} \\ 
		\hline
		& Ana & Per &  Ana &Per & Ana &Per & Ana &Per  \\
		\hline 
		Gaussian  & 2.87 & 2.88 &2.93 & 2.95 & 3.00 & 3.02 & 3.11 & 3.12  \\
		$d = 100$ & 2.86 & 2.86 & 2.93 & 2.94 & 3.00 & 3.03 & 3.10 & 3.08 \\
		Gaussian & 2.82 & 2.83 & 2.88 & 2.89 & 2.94 & 2.93 & 3.04 & 3.02  \\
		$d = 500$ & 2.82 & 2.78 & 2.88 & 2.88 & 2.94 & 2.94  & 3.04 & 3.04 \\ 
		Gaussian & 2.81 & 2.79 & 2.87 & 2.84 & 2.94 & 2.93 & 3.04 & 3.04  \\
		$d = 1000$  & 2.81 & 2.78 & 2.87 & 2.87  & 2.94 & 2.93 & 3.03 & 3.00 \\
		\hline
		MV-$t_{5}$ & 2.88 & 2.91  & 2.94 & 2.93  & 3.02 & 3.05 & 3.13 & 3.14  \\ 
		$d = 100$& 2.88 & 2.92 & 2.94 & 2.97  & 3.02 & 3.03 & 3.13 & 3.13 \\ 
		MV-$t_{5}$ & 2.81 & 2.82  & 2.86 & 2.86 & 2.94& 2.93  & 3.04 & 3.04 \\
		$d = 500$ & 2.81 & 2.80  & 2.87 & 2.86  & 2.93 & 2.92  & 3.03 & 3.02\\ 
		MV-$t_{5}$& 2.79 & 2.79  & 2.86 & 2.86 & 2.92 & 2.90  & 3.01 & 3.00  \\
		$d = 1000$ & 2.79 & 2.79 & 2.85 & 2.85  & 2.91 & 2.91   & 3.01 & 3.01 \\
		\hline 
		Log-normal  & 3.01 & 3.14 & 3.08 & 3.22 & 3.18 & 3.29 & 3.31 & 3.48  \\
		$d = 100$& 3.01 & 3.12  & 3.08 & 3.23 & 3.18 & 3.30  & 3.32 & 3.49 \\
		Log-normal  & 2.91 & 2.97   & 2.98 & 3.05  & 3.06 & 3.11& 3.18 & 3.24 \\
		$d = 500$ & 2.90 & 2.96  & 2.97 & 3.04  & 3.05 & 3.09  & 3.17 & 3.24\\ 
		Log-normal  & 2.88 & 2.90 & 2.94 & 2.97 & 3.02 & 3.04 & 3.13 & 3.17  \\
		$d = 1000$ & 2.88 & 2.92  & 2.94 & 2.98 & 3.02 & 3.07   & 3.13 & 3.17 \\
		\hline
	\end{tabular} 
\end{table} 
\begin{table}[h!]
	\centering 
	\caption{Critical values for the single change-point scan statistic $\max_{n_0 \le t \le n_1} Z_{W,0.8}(t)$ at 0.05 significance level}
	\label{tab:zwcv2}
	\begin{tabular}{c|cc|cc|cc|cc} 
		\hline  
		& \multicolumn{2}{c|}{$n_0 = 100$} &  \multicolumn{2}{c|}{$n_0 = 75$} &  \multicolumn{2}{c|}{$n_0 = 50$} &  \multicolumn{2}{c}{$n_0 = 25$}  \\ 
		\hline
		& Ana & Per & Ana &Per &Ana &Per & Ana &Per  \\
		\hline 
		Gaussian & 2.84 & 2.84 & 2.89 & 2.88 & 2.97 & 2.98 &3.07 & 3.04  \\
		$d = 100$ &  2.84 & 2.84 &  2.90 & 2.90 & 2.97 & 2.99 &  3.07 & 3.06 \\  
		Gaussian &  2.80 & 2.80 & 2.85 & 2.84 & 2.92 & 2.91 & 3.01 & 2.99  \\
		$d = 500$ & 2.80 & 2.80 & 2.85 & 2.87 & 2.92 & 2.92 & 3.01 & 3.00 \\ 
		Gaussian &2.79 & 2.79 & 2.85 & 2.85 & 2.92 & 2.89  & 3.01 & 2.96  \\
		$d = 1000$ &  2.80 & 2.80 & 2.85 & 2.82 & 2.91 & 2.91  & 3.00 & 2.99  \\
		\hline 
		MV-$t_{5}$ & 2.85 & 2.88 & 2.91 & 2.94 & 2.99 & 2.99  & 3.09 & 3.11  \\
		$d = 100$ & 2.85 & 2.87  & 2.91 & 2.95 & 2.99 & 3.00& 3.09 & 3.06 \\  
		MV-$t_{5}$ & 2.79 & 2.79 & 2.85 & 2.84  & 2.92 & 2.92 & 3.01&3.01\\
		$d = 500$ &  2.80 & 2.80  & 2.85 & 2.83 & 2.92& 2.92  & 3.01 & 3.00 \\  
		MV-$t_{5}$ & 2.79 & 2.79  & 2.85 & 2.84 & 2.92 & 2.92 & 3.00 & 3.01  \\
		$d = 1000$  & 2.80 & 2.80 & 2.85 & 2.83 & 2.92 & 2.92& 2.99 & 2.99  \\
		\hline 
		Log-normal &  2.96 & 3.04  & 3.04 & 3.11  & 3.13 & 3.24  & 3.25 & 3.39  \\
		$d = 100$ & 2.96 & 3.05 &  3.04 & 3.12 & 3.13 & 3.21 & 3.26 & 3.40 \\ 
		Log-normal & 2.88 & 2.92 & 2.94 & 2.96  & 3.02 & 3.04  & 3.13&3.19\\
		$d = 500$  & 2.88 & 2.93  & 2.94 & 2.97 & 3.02 & 3.06  & 3.13 & 3.19 \\ 
		Log-normal  & 2.84 & 2.84  & 2.92 & 2.97 & 2.99 & 3.01 & 3.10 & 3.12  \\
		$d = 1000$  & 2.85 & 2.86  & 2.92 & 2.94 & 2.99 & 3.01  & 3.10 & 3.13  \\
		\hline 
	\end{tabular} 
\end{table} 

Table \ref{tab:zdcv} shows the results of the scan statistic of $Z_{D}(t)$. `Ana' denotes the skewness-corrected analytical critical values and `Per' presents the critical values obtained from 10,000 permutation. We see that the analytical $p$-value approximation works well in all cases.

Table \ref{tab:zwcv1} shows the results of the scan statistic of $Z_{W,1.2}(t)$, and Table \ref{tab:zwcv2} shows the results of the scan statistic of $Z_{W,0.8}(t)$.  We see that the skewness-corrected analytical $p$-value approximations also work well for both scan statistics.  The choice of $r$ is discussed in Section \ref{subsec:fast}.



\subsection{Fast tests} \label{subsec:fast}

For $\{Z_{W,r}(t)\}$ ($r\ne 1$), when $r$ is close to 1, it converges to the Gaussian process is slow.  On the other hand, the performance of the test decreases as $r$ goes away from 1 for location alternatives. Table \ref{table:zw} shows the estimated power of $Z_{W,r}(t)$ under various $r$ for Gaussian data where the first 100 observations are generated from from $N_{d}(\mu_{1}, I_{d})$ and the second 100 observations are generated from $N_{d}(\mu_{2}, I_{d})$, where $\Delta = \|\mu_{1}-\mu_{2}\|_{2}$. The significance level is set to be 0.05 and the $p$-values of each test are approximated by 10,000 permutations for fair comparison. 

\begin{table} [h]
	\caption{Estimated power (by 100 simulation runs) of $Z_{W,r}(t)$ at 0.05 significance level}
	\label{table:zw}
	\centering
	\begin{tabular}{c|c|c|c|c|c}
		\hline
		\multicolumn{6}{c}{Location Alternatives}  \\
		\hline
		$d$ & 10 & 30 & 50 & 70 & 100  \\
		$\Delta$ & 0.47 & 0.60 & 0.77 & 0.96 & 1.13 \\  
		\hline
		$r=1.3$ & 0.23 & 0.16 & 0.21 & 0.24 & 0.31   \\
		$r=1.2$ & 0.32 & 0.22 & 0.34 & 0.40 & 0.47   \\
		$r=1.1$ & 0.36 & 0.33 & 0.46 & 0.67 & 0.72   \\
		$r=1.0$ & \textbf{0.42} & \textbf{0.43} & \textbf{0.56} & \textbf{0.80} & \textbf{0.88}  \\
		$r=0.9$ & 0.35 & 0.34 & 0.45 & 0.66 & 0.80   \\
		$r=0.8$ & 0.21 & 0.25 & 0.27 & 0.38 & 0.49   \\
		$r=0.7$ & 0.14 & 0.14 & 0.16 & 0.20 & 0.26  \\
		\hline
	\end{tabular}
\end{table}

To make use of the asymptotic result of $Z_{W,r}(t)$ and maximize the power of the test, we propose to use $Z_{W,1.2}(t)$ and $Z_{W,0.8}(t)$ together. The power of the test could be enhanced in some common scenarios by using both test statistics together as they cover different regions of alternatives and their $r$'s are far enough away from 1 so that the asymptotic results kick in while maintaing a good power.

We now define two fast tests based on the asymptotic results. Let $p_{D}$ , $p_{W,1.2}$, and $p_{W,0.8}$ be the approximated $p$-values of the test that reject for large values of $|Z_{D}(t)|$, $Z_{W,1.2}(t)$, and $Z_{W,0.8}(t)$, respectively. 
\begin{itemize}
	\item $\textrm{fGKCP}_{1}$: rejects the null hypothesis of homogeneity if $3\min(p_{D}, p_{W,1.2}, p_{W,0.8})$ is less than the significance level.
	\item $\textrm{fGKCP}_{2}$: rejects the null hypothesis of homogeneity if $2\min(p_{W,1.2}, p_{W,0.8})$ is less than the significance level.
\end{itemize}
It is expected that $\textrm{fGKCP}_{1}$ performs well for a wide range of alternatives, especially for scale alternatives due to $Z_{D}(t)$ (reasoning goes back to Figure \ref{fig:motivation}). Since $Z_{W}(t)$ is sensitive to location alternatives, we expect $\textrm{fGKCP}_{2}$ to be powerful for location alternatives. Furthermore, according to the simulation results in Section \ref{sec:performance}, it turns out that $\textrm{fGKCP}_{2}$ can also detect variance changes to some extent as $r = 1.2, 0.8$ cover more types of alternatives than $r=1$. When the null hypothesis is rejected based on $\textrm{fGKCP}_{1}$ or $\textrm{fGKCP}_{2}$, the location of change-point can be estimated by $\max_{n_{0}\le t \le n_{1}}\textrm{GKCP}(t)$.

\begin{remark}
	We adopt the Bonferroni procedure for the fast tests to combine the advantages of each test statistic. To improve the power of the tests, the Simes procedure might be used and this also controls type I error well empirically (see Section \ref{sec:conclusion}). 
\end{remark}


\section{Performance of the new tests} \label{sec:performance}

We examine the performance of the new tests under various simulation settings. Each data sequence in the simulation is of length $n=200$ with various dimensions $d$, where $y_{1},\ldots,y_{\tau} \stackrel{iid}{\sim} F_{0}$ and $y_{\tau+1},\ldots,y_{n} \stackrel{iid}{\sim} F_{1}$. Here, $\tau$ is the change-point. We consider the following setting:
\begin{itemize}
	\item Multivariate Gaussian data Type I: $F_{0}\sim N_{d}(\textbf{0}_{d},\Sigma)$ vs.  $F_{1}\sim N_{d}(a\textbf{1}_{d},\sigma^2\Sigma)$, where $\Delta = \|a\textbf{1}_{d}\|_{2}$ and $\Sigma_{i,j} = 0.4^{|i-j|}$.
	\item Multivariate Gaussian data Type II: $F_{0}\sim N_{d}(\textbf{0}_{d},\Sigma)$ vs.  $F_{1}\sim N_{d}(a\nu_{d},\sigma^2\Sigma)$, where $\Delta = \|a\nu_{d}\|_{2}$, $d$-dimensional vector $\nu_{d}$ with half of it being zeros and half of it being 1's, and $\Sigma_{i,j} = 0.4^{|i-j|}$.
	\item Chi-square data: $\Sigma^{1/2}u_1$ vs.  $(\sigma^2\Sigma)^{1/2}u_2+ a\textbf{1}_{d}$, 
	where $\Delta = \|a\textbf{1}_{d}\|_{2}$ and $u_1$ and $u_2$ are length-$d$ vectors with each component i.i.d. from the $\chi_{3}^2$ distribution. 
	\item Multivariate log-normal data: $F_{0}\sim \exp(N_{d}(\textbf{0}_{d},\Sigma))$ vs.  $F_{1}\sim \exp(N_{d}(a\textbf{1}_{d},\Sigma))$, where $\Delta = \|a\textbf{1}_{d}\|_{2}$ and $\Sigma_{i,j} = 0.4^{|i-j|}$.
\end{itemize}
We simulate 100 datasets to estimate the power of the tests and the significance level is set to be 0.05 for all tests. To examine the empirical size of the test, we simulate 1,000 datasets. We also examine the accuracy of the estimated change-point location and the count where the location of estimated change-point is within 20 from the true change-point when the null is rejected. 

It is usually hard to offer false positive controls as well as the estimation of the location of change-points. We compare the results for the new tests to the recent feasible kernel-based method, KCP \cite{arlot2019kernel}, which can be implemented by an $\texttt{R}$ package $\texttt{ecp}$ \cite{james2013ecp}. We also compare the new tests with other feasible nonparametric methods using interpoint distances (ECP) \cite{matteson2014nonparametric} and similarity graphs (GCP) \cite{chu2019asymptotic}, which was implemented by $\texttt{R}$ packages $\texttt{ecp}$ and $\texttt{gSeg}$, respectively. Here, we approximate the $p$-value by 1,000 permutation for GKCP and ECP, and use the max-type method with 5-MST for GCP, following the suggestion in \cite{chen2015graph}. Lastly, we include the method using Fr$\acute{\textrm{e}}$chet means and variances (FCP) with the $p$-value approximated by 5,000 bootstrap replicates  \cite{dubey2020frechet}.

\begin{table} [h!]
	\caption{The number of null rejection, out of 100, and the number of accurately detected change-points (in parentheses) for multivariate Gaussian data Type I.  The numbers larger than 95\% of the largest one in each scenario are in bold}
	\label{tab:simul:normal}
	\centering
	\begin{tabular}{c|cccc}
		\hline
		& \multicolumn{4}{c}{Mean Change ($\tau$ at center)}\\
		\hline
		$d$ & 100 & 500 & 1000  & 2000 \\ 
		$\Delta$   & 1.20 & 1.90 & 2.40 & 3.13  \\
		\hline
		$\textrm{fGKCP}_{1}$ & 50 (43)& 68 (62)& 78 (76) & \textbf{96 (95)} \\ 
		$\textrm{fGKCP}_{2}$ & 58 (49)& 73 (67)& 84 (80) & \textbf{97 (96)}\\
		GKCP  & \textbf{75 (63)} &  \textbf{88 (82)} &  \textbf{95 (91)} &  \textbf{99 (98)}  \\
		KCP &  71 (61) &  \textbf{85 (79)} &  \textbf{93 (90)} &  \textbf{98 (97)}   \\
		ECP &  \textbf{76 (65)} &  \textbf{89 (79)} &  \textbf{96 (90)} &  \textbf{99 (95)} \\
		GCP & 22 (9) & 27 (14) & 34 (20) & 46 (32)  \\
		FCP & 6 (1) & 1 (0) & 0 (0) & 0 (0)   \\
		\hline
	\end{tabular} 
	
	\vspace*{0.3cm}  
	
	\begin{tabular}{c|cccc}
		\hline
		& \multicolumn{4}{c}{Variance Change ($\tau$ at center)}  \\ 
		\hline
		$d$ & 100 & 500 & 1000  & 2000\\ 
		$\sigma^2$   & 1.07 & 1.04 & 1.03 & 1.0\\
		\hline
		$\textrm{fGKCP}_{1}$ &  \textbf{46 (30)}&  \textbf{68 (52)} &  \textbf{79 (64)} &  \textbf{93 (81)}\\
		$\textrm{fGKCP}_{2}$ &  40 (25)&  58 (43)&  68 (54) &  85 (73)\\
		GKCP  &  41 (27) &  \textbf{67 (51)} &  \textbf{79 (63)} &  \textbf{93 (80)} \\
		KCP & 18 (2) & 15 (3) & 12 (2) & 7 (1)  \\
		ECP & 5 (2) & 6 (2) & 6 (2) & 6 (2)  \\
		GCP & 27 (11) & 40 (21) & 49 (27) & 64 (41) \\
		FCP & 13 (5) & 0 (0) & 0 (0) & 0 (0)  \\
		\hline
	\end{tabular} 
\end{table}
\begin{table}[h!]
	\caption{Results for multivariate Gaussian data Type II}
	\label{tab:simul:normal2}
	\centering
	\begin{tabular}{c|cccc}
		\hline
		& \multicolumn{4}{c}{Mean Change ($\tau$ at center)} \\ 
		\hline
		$d$ & 100 & 500 & 1000  & 2000 \\ 
		$\Delta$   & 0.99 & 2.37 & 2.46 & 3.16 \\
		\hline
		$\textrm{fGKCP}_{1}$ & 17 (10)& 39 (31)& 57 (51) & 84 (81) \\ 
		$\textrm{fGKCP}_{2}$ & 21 (13)& 46 (36)& 64 (57) & 89 (87) \\
		GKCP  &  \textbf{34 (24)} &  \textbf{64 (52)} &  \textbf{81 (72)} &  \textbf{97 (94)} \\
		KCP & 31 (22) &  59 (49) & 78 (70) &  \textbf{94 (92)}   \\
		ECP & 32 (21) &  \textbf{63 (52)} &  \textbf{85 (75)} &  \textbf{98 (90)}  \\
		GCP & 12 (3) & 18 (6) & 24 (11) & 31 (20) \\
		FCP & 4 (0) & 0 (0) & 0 (0) & 0 (0)   \\
		\hline
	\end{tabular} 
	
	\vspace*{0.3cm}  
	
	\begin{tabular}{c|cccc}
		\hline
		& \multicolumn{4}{c}{Mean and Variance Change ($\tau$ at center)} \\ 
		\hline
		$d$ & 100 & 500 & 1000  & 2000 \\
		$\Delta$ & 0.65 & 0.69 & 0.70  & 0.71  \\ 
		$\sigma^2$   & 1.06 & 1.04 & 1.03 & 1.03  \\
		\hline
		$\textrm{fGKCP}_{1}$ &  \textbf{46 (30)} &  \textbf{63 (46)} &  \textbf{79 (63)} &  \textbf{99 (90)}  \\
		$\textrm{fGKCP}_{2}$ & 43 (27) & 58 (43) & 72 (56) &  \textbf{96 (88)} \\
		GKCP  & 42 (27) &  \textbf{61 (45)} &  \textbf{78 (61)} & 90 (90) \\
		KCP & 5 (2) & 2 (1) & 1 (0) & 1 (0)  \\
		ECP & 11 (5) & 7 (3) & 5 (2) & 7 (2)  \\
		GCP & 23 (9) & 34 (17) & 47 (27) & 77 (54)  \\
		FCP & 12 (6) & 0 (0) & 0 (0) & 0 (0)\\
		\hline
	\end{tabular} 
\end{table}
\begin{table} [h!]
	\caption{Results for chi-square data}
	\label{tab:simul:chi}
	\centering
	\begin{tabular}{c|cccc}
		\hline
		& \multicolumn{4}{c}{Mean Change ($\tau$ at center)} \\
		\hline
		$d$ & 100 & 500 & 1000  & 2000\\ 
		$\Delta$   & 2.60 & 4.24 & 5.69 & 8.04 \\
		\hline
		$\textrm{fGKCP}_{1}$ & 24 (16)& 40 (35)& 45 (43) & 80 (79)\\ 
		$\textrm{fGKCP}_{2}$ & 29 (19)& 46 (40)& 63 (60) & 87 (86) \\
		GKCP  & \textbf{51 (40)} &  \textbf{74 (64)} &  \textbf{94 (88)} &  \textbf{99 (99)} \\
		KCP &  4 (0) &   4 (0) &   4 (0) &  3 (0)\\
		ECP &  \textbf{58 (45)} &  \textbf{78 (66)} &  \textbf{94 (87)} &  \textbf{99 (96)} \\
		GCP & 22 (8) & 26 (10) & 32 (19) & 54 (38) \\
		FCP & 5 (0) & 2 (0) & 0 (0) & 0 (0)  \\
		\hline
	\end{tabular} 
	
	\vspace*{0.3cm}  
	
	\begin{tabular}{c|cccc}
		\hline
		& \multicolumn{4}{c}{Variance Change ($\tau$ at center)}  \\ 
		\hline
		$d$ & 100 & 500 & 1000  & 2000 \\ 
		$\sigma^2$   & 1.23 & 1.11 & 1.10 & 1.09\\
		\hline
		$\textrm{fGKCP}_{1}$ &  \textbf{78 (62)}&  \textbf{76 (56)} &  \textbf{95 (82)} &  \textbf{99 (92)}  \\
		$\textrm{fGKCP}_{2}$ &  \textbf{81 (64)}&  \textbf{78 (57)}&  \textbf{95 (83)} &  \textbf{99 (92)} \\
		GKCP  &  75 (60) &  71 (53) &  \textbf{92(80)} &  \textbf{99 (92)}  \\
		KCP & 20 (16) & 6 (5) & 10 (9) & 5 (4) \\
		ECP & 59 (46) & 30 (18) & 37 (26) & 48 (38)  \\
		GCP & 27 (11) & 29 (10) & 44 (22) & 63 (37) \\
		FCP & 53 (35) & 6 (2) & 2 (0) & 0 (0) \\
		\hline
	\end{tabular} 
\end{table}
\begin{table} [h!]
	\caption{Results for multivariate log-normal data}
	\label{tab:simul:lognormal}
	\centering
	\begin{tabular}{c|cccc}
		\hline
		& \multicolumn{4}{c}{Mean Change ($\tau$ at center)}  \\ 
		\hline
		$d$ & 100 & 500 & 1000  & 2000 \\ 
		$\Delta$   & 1.20 & 1.90 & 2.30 & 3.04  \\
		\hline
		$\textrm{fGKCP}_{1}$ & 47 (35)& 70 (57)& 81 (71) & \textbf{96 (90)} \\
		$\textrm{fGKCP}_{2}$ & 55 (41) &  76 (63) & 85 (75) &  \textbf{97 (91)} \\
		GKCP  & 63 (48) &  \textbf{83 (68)} &  \textbf{91 (80)} &  \textbf{99 (93)} \\
		KCP & 20 (16) & 6 (5) & 10 (9) & 5 (4)  \\
		ECP &  \textbf{69 (52)} &  \textbf{85 (72)} &  \textbf{91 (80)} &  \textbf{98 (91)}   \\
		GCP & 32 (12) & 33 (7) & 32 (6) & 36 (8)  \\
		FCP & 32 (18) & 57 (40) & 69 (53) & 83 (70)  \\
		\hline
	\end{tabular} 
\end{table}
\begin{table} [h!]
	\caption{Empirical size of the tests at 0.05 significance level}
	\label{tab:simul:size}
	\centering
	\begin{tabular}{c|cccc}
		\hline
		& \multicolumn{4}{c}{Multivariate Gaussian}\\ 
		\hline
		$d$ & 100 & 500 & 1000  & 2000 \\
		\hline
		$\textrm{fGKCP}_{1}$ & 0.032 & 0.047 & 0.047 & 0.037  \\
		$\textrm{fGKCP}_{2}$ & 0.043 & 0.057 & 0.055 & 0.052  \\
		GKCP  & 0.052 & 0.049 & 0.053 & 0.049 \\
		KCP & 0.067 & 0.045 & 0.060 & 0.040   \\
		ECP & 0.054 & 0.043 & 0.056 & 0.045   \\
		GCP & 0.072 & 0.073 & 0.069 & 0.077 \\
		FCP & 0.018 & 0.001 & 0.000 & 0.000  \\
		\hline
	\end{tabular} 

   \vspace*{0.3cm}  

    \begin{tabular}{c|cccc}
    	\hline
        & \multicolumn{4}{c}{Multivariate log-normal} \\ 
    	\hline
    	$d$ & 100 & 500 & 1000  & 2000 \\
    	\hline
    	$\textrm{fGKCP}_{1}$ & 0.038 & 0.039 & 0.041 & 0.036 \\
    	$\textrm{fGKCP}_{2}$ & 0.051 & 0.050 & 0.050 & 0.055 \\
    	GKCP & 0.049 & 0.051 & 0.038 & 0.056 \\
    	KCP & 0.093 & 0.040  & 0.081 & 0.067  \\
    	ECP & 0.054 & 0.057 & 0.051 & 0.042  \\
    	GCP & 0.090 & 0.132 & 0.098 & 0.113 \\
    	FCP & 0.053 & 0.051 & 0.036 & 0.027  \\
    	\hline
    \end{tabular} 
\end{table}

Table \ref{tab:simul:normal} and \ref{tab:simul:normal2} show the number of rejection for the multivariate Gaussian data with different means and$\slash$or variances. The count where the estimated change-point is within 20 from the true change-point is provided in parentheses when the null hypothesis is rejected. We see that KCP and ECP perform well for location alternatives, while they have considerable low or no power for scale alternatives. On the other hand, the new test GKCP performs very well for both location and scale alternatives, and the fast tests, $\textrm{fGKCP}_{1}$ and $\textrm{fGKCP}_{2}$, also perform well. Other tests, GCP and FCP, do not work well for Gaussian settings. 

Table \ref{tab:simul:chi} shows results for the chi-square data. We see that KCP has no power (the penalty constant is difficult to optimize for this dataset). ECP still performs well for location alternatives, but it loses power for scale alternatives. On the other hand, the new tests in general perform well for both loacation and scale alternatives. GCP and FCP exhibit no or lower power than the new tests.

Table \ref{tab:simul:lognormal} shows results for the multivariate log normal data. Here, alternatives yield the changes in both the mean and variance of distributions. We still see that the new tests exhibit high power not only for symmetric distributions but also for asymmetric distributions under moderate to high dimensions. However, KCP and GCP lose power in this case, while ECP still performs well. Compared with Gaussian settings, FCP exhibits high power, but it is outperformed by the new tests.

The empirical size of the tests at 0.05 significance level for the multivariate Gaussian and log-normal data is presented in Table \ref{tab:simul:size}. We see that the new tests control the type I error rate well. However, KCP relies on a cumbersome method, such as the line search, to find the suitable penalty constant and this step is very sensitive, so it is difficult to control the type I error well.

\begin{table}[h!]
	\caption{Average runtimes in seconds from 10 simulations for each length $n$. All experiments were run by $\texttt{R}$ on 2.2 GHz Intel Core i7}
	\label{tab:simul:time}
	\centering
	\begin{tabular}{crrrrrr}
		\hline
		$n$ &200 & 400 & 600 & 800 & 1000 & 2000 \\ 
		\hline
		$\textrm{fGKCP}_{1}$ & 0.04 & 0.24 & 0.72 & 1.75 & 3.33 & 25.58 \\
		$\textrm{fGKCP}_{2}$ & 0.04 & 0.22 & 0.71 & 1.71 & 3.29 & 25.44 \\
		GKCP & 4.63 & 8.51 & 17.66 & 32.45 & 49.95 & 201.50 \\
		KCP & 0.21 & 3.28 & 17.27 & 53.32 & 132.00 & 2161.83 \\
		ECP & 1.44 & 5.05 & 12.00 & 19.22 & 30.25 & 144.38 \\
		GCP & 0.05 & 0.13 & 0.27 & 0.34 & 0.59 & 2.02 \\
		FCP & 26.57 & 94.37 & 209.10 & 369.50 & 544.00 & 2251.70 \\
		\hline
	\end{tabular}
\end{table}

We also compare the computational cost of the tests and check runtimes of the tests for Gaussian data under various $n$. Table \ref{tab:simul:time} shows average runtimes for each length $n$ when $d=100$. The new methods are implemented in an $\texttt{R}$ package $\texttt{kerSeg}$. We use 1,000 permutations for GKCP and ECP. We first see that $\textrm{fGKCP}_{1}$ and $\textrm{fGKCP}_{2}$ are faster than GKCP since GKCP relies on the permutation approach. KCP is fast when the sequence is short, but its running time increases dramatically in $n$.  When $n=2,000$, it needs 36 minutes on average. Note that the average runtimes of KCP in Table \ref{tab:simul:time} only present the actual testing runtimes. If we consider the runtime for choosing the tuning parameter, which is essential for KCP according to Table \ref{tab:simul:tuning}, KCP is almost  computationally infeasible to run. FCP is as slow as KCP. ECP relies on the permutation approach, so it is slower than the fast tests. GCP is the fastest among the tests in the comparison.

Overall, simulation results shows that the new tests exhibit high power for a wide range of alternatives. Unlike the existing kernel change-point detection method, the new tests are effective and easy to implement without any time-consuming procedures, such as parameter tuning, as long as the kernel matrix is computed. In practice, $\textrm{fGKCP}_{1}$ and $\textrm{fGKCP}_{2}$ would be preferred as they are faster than GKCP. However, if the test result is ambiguous and further investigation is needed, such as $p$-value close to the nominal level, the permuation test of GKCP would also be useful.


\section{A real data example} \label{sec:real}

We apply the new tests to a phone-call network dataset, collected by the MIT Media Laboratory. The study involved 87 subjects who used mobile phones with a pre-installed device that can record call logs. The study lasted for 330 days from July 2004 to June 2005 \cite{eagle2009inferring}. We use it to illustrate the new tests by detecting any change in the phone-call pattern among subjects over time. This can be viewed as the change of friendship along time.

We bin the phone-calls by day and in total construct $n=330$ of networks with 87 subjects as nodes. We encode each network by the adjacency matrix with value 1 for element $(i,j)$ if subject $i$ called $j$ on day $t$ and 0 otherwise. We then construct the Gaussian kernel matrix with the median heuristic using the vectorized adjacency matrix.

We apply the single change-point detection method to the phone-call network dataset recursively in order to detect all possible change-points. Since this dataset has a lot of noise, we focus on the estimated change-points with $p$-value less than 0.001. 

\begin{table}[h!]
	\caption{Estimated change-points}
	\label{real:phone}
	\centering
	\begin{tabular}{c|ccccccc}
		\hline
		& \multicolumn{5}{c}{Days (t)} \\ 
		\hline
		Estimated change-points & 53 & 90 & 141 & 251 & 293  \\
		\hline
	\end{tabular}
\end{table}

\begin{figure}[h]
	\centering
	\includegraphics[width=3in]{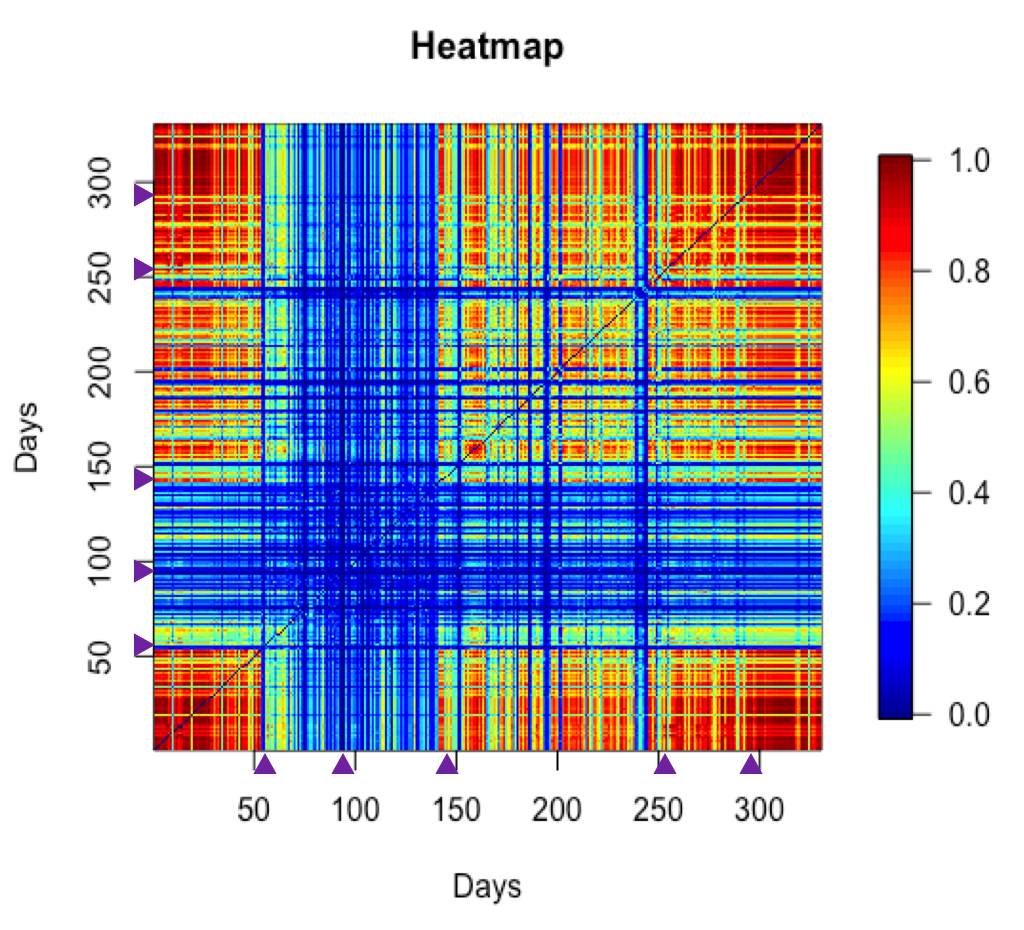} 
	\caption{The heatmap of the kernel matrix corresponding to 330 networks. Purple triangles in the heatmap indicate estimated change-points by the new tests.} 
	\label{fig:heatmap}
\end{figure}

Table \ref{real:phone} shows the estimated change-points until the new tests do not reject the null. In this analysis, all new tests (GKCP, $\textrm{fGKCP}_{1}$, $\textrm{fGKCP}_{2}$) yield the same results. Since the underlying distribution of the dataset is unknown, we perform a sanity check on the kernel matrix of the whole period (Figure \ref{fig:heatmap}). It is evident that there are some changes occuring in the period and they match the results of the new test fairly well.

\begin{table}[h!]
	\caption{Estimated change-points and nearby academic events. The dates of the academic events are from the 2011-2012 academic calendar of MIT that is the closest academic calendar of MIT to 2004-2005 available online}
	\label{tab:event}
	\centering
	\begin{tabular}{c|c}
		\hline
		Estimated change-points & Nearby academic events \\ 
		\hline
		$n=53$: 2004/09/10 & 2004/09/07: Fall classes begin \\ 
		$n=90$: 2004/10/17 & 2004/10/14: Family weekend \\
		$n=141$: 2004/12/07 &  2004/12/14:  Last day of Fall classes \\
		$n=251$: 2005/03/27 & 2005/03/26: Spring break begins \\
		$n=293$: 2005/05/08 & 2005/05/17: Last day of Spring classes \\
		\hline
	\end{tabular}
\end{table}

We also compare the results of the new tests with their nearby academic events (Table \ref{tab:event}). We see that the new tests detect change-points at around the beginning of the Fall term, family weekend, and the end of the Fall term that could cause phone-call pattern changes among subjects. The new tests also detect the Spring break and the end of the Spring term. These are all reasonable times when there are some significant changes in phone-call pattern.


\section{Discussion and conclusion} \label{sec:conclusion}

We proposed the new kernel-based scan statistic, GKCP, for the testing and estimation of change-points. The new tests are versatile and effective for a wide range of alternatives. We also proposed two fast tests, $\textrm{fGKCP}_{1}$ and $\textrm{fGKCP}_{2}$, that have analyitic $p$-value approximations. The new tests exhibit superior power and work well particularly for high-dimensional settings. In practice, we recommend to use $\textrm{fGKCP}_{1}$ and $\textrm{fGKCP}_{2}$ as they are fast to implement. When the results are ambiguous, the permutation test based on GKCP could be run for the final conclusion.

Since the Bonferroni procedure is a bit conservative, the Simes procedure may be used to improve the power of the fast tests ($\textrm{fGKCP}_{1}$, $\textrm{fGKCP}_{2}$). Let $p_{(1)} \le p_{(2)} \le p_{(3)}$ be the ordered $p$-values of $p_{D}$, $p_{W,1.2}$, and $p_{W,0.8}$ and $p_{(1)}' \le p_{(2)}'$ be the ordered $p$-values of $p_{W,1.2}$ and $p_{W,0.8}$. Then, the fast tests are defined that
\begin{itemize}
	\item $\textrm{fGKCP}_{1}$-Simes: rejects the null hypothesis of homogeneity if $\min(3p_{(1)}, 1.5p_{(2)}, p_{(3)})$ is less than the significance level.
	\item $\textrm{fGKCP}_{2}$-Simes: rejects the null hypothesis of homogeneity if $\min(2p_{(1)}', p_{(2)}')$ is less than the significance level.
\end{itemize}
It has been shown that the Simes procedure is exact under independent distributions, while it becomes conservative under positively dependent distributions and slightly conservative under negatively dependency. There have been a lot of works to prove the validity of the Simes test under dependency \cite{block1982some, hochberg1995extensions, samuel1996simes, sarkar1997simes, block2008negative, finner2017simes, gou2018hochberg}, but they are restricted to special cases. Nevertheless, the Simes test is widely used in many applications. \cite{rodland2006simes} proved that the overall relative deviation of the Simes $p$-value from the true $p$-value is strongly bounded and showed that, although the Simes procedure may be liberal, it cannot be consistently. It is therefore reasonably expected that the Simes $p$-value will be asymptotically valid in most practical cases.
\begin{table} [h!]
	\caption{Empirical size of the tests at 0.05 significance level}
	\label{tab:bon}
	\centering
	\begin{tabular}{c|cccc}
		\hline
		& \multicolumn{4}{c}{Multivariate Gaussian} \\ 
		\hline
		$d$ & 100 & 500 & 1000  & 2000 \\
		\hline
		$\textrm{fGKCP}_{1}$ & 0.032 & 0.047 & 0.047 & 0.037\\ 
		$\textrm{fGKCP}_{1}$-Simes & 0.036 & 0.048 & 0.048 & 0.038  \\ 
		$\textrm{fGKCP}_{2}$ & 0.043 & 0.057 & 0.055 & 0.052  \\
		$\textrm{fGKCP}_{2}$-Simes & 0.044 & 0.057 & 0.057 & 0.052  \\
		\hline
	\end{tabular} 

    \vspace*{0.3cm}  

    \begin{tabular}{c|cccc}
    	\hline
    	& \multicolumn{4}{c}{Multivariate log-normal} \\ 
    	\hline
    	$d$ & 100 & 500 & 1000  & 2000\\
    	\hline
    	$\textrm{fGKCP}_{1}$ & 0.030 & 0.039 & 0.036 & 0.035 \\ 
    	$\textrm{fGKCP}_{1}$-Simes &  0.031 & 0.040 & 0.036 & 0.036 \\ 
    	$\textrm{fGKCP}_{2}$ &  0.051 & 0.052 & 0.055 & 0.052 \\
    	$\textrm{fGKCP}_{2}$-Simes &  0.051 & 0.052 & 0.055 & 0.052 \\
    	\hline
    \end{tabular} 
\end{table} 
Table \ref{tab:bon} shows the empirical size of the tests for the multivariate Gaussian and log-normal data used in Section \ref{sec:performance}. We see that the Simes procedure also controls type I error well. Hence, if we want to focus on the performance of the test and improve the power of the fast tests, the Simes procedure would be useful for the fast tests.

The new methods detect the most significant single change-point or changed-interval in the sequence. If two or more changes are presented in the sequence, the new methods can be applied recursively with multiple change-point detection techniques, such as binary segmentation, circular binary segmentation, or wild binary segmentation \cite{vostrikova1981detecting, olshen2004circular, fryzlewicz2014wild}.

\ifCLASSOPTIONcaptionsoff
  \newpage
\fi



%

\bibliographystyle{IEEEtran}
\bibliography{Bibliography-MM-MC}

%








\end{document}